\documentclass[12pt]{article}
\usepackage{graphicx,amssymb}
\usepackage[figuresright]{rotating}

\title{The  static force in  background  perturbation theory}
\author{  A.M.Badalian and A.I.Veselov \\
State Research Center, Institute of Theoretical and\\ Experimental
Physics, Moscow, 117218 Russia\\ }

\newcommand{\beq}{\begin{eqnarray}}
 \newcommand{\eeq}{\end{eqnarray}}
\newcommand{\be}{\begin{equation}}
 \newcommand{\ee}{\end{equation}}

 \def\la{\mathrel{\mathpalette\fun <}}
\def\ga{\mathrel{\mathpalette\fun >}}
\def\fun#1#2{\lower3.6pt\vbox{\baselineskip0pt\lineskip.9pt
\ialign{$\mathsurround=0pt#1\hfil ##\hfil$\crcr#2\crcr\sim\crcr}}}

\newcommand{{\SD}}{\rm SD}

\newcommand{\lan}{\langle}
\newcommand{\ran}{\rangle}

\begin{document}

\maketitle
\begin{abstract}
The static force $F_B(r)$ and the strong coupling $\alpha_F(r)$,
which defines the gluon-exchange part of $F_B(r)$, are studied in
QCD background perturbation theory (BPT). In the region $r\la 0.6
$ fm $\alpha_F(r)$ turns out to be essentially smaller than the
coupling $\alpha_B(r)$ in the static potential. For the
dimensionless function $\Phi_B(r) = r^2 F_B(r)$ the characteristic
values $\Phi_B(r_1) =1.0$ and $\Phi_B(r_0)=1.65$ are shown to be
reached at the   following $Q\bar Q$ separations:
$r_1\sqrt{\sigma} =0.77, ~~ r_0\sqrt{\sigma} =1.09$ in quenched
approximation and  $r_1\sqrt{\sigma}=0.72,~~
r_0\sqrt{\sigma}=1.04$ for $n_f =3$. The numbers obtained appear
to be by only 8\% smaller than those calculated in lattice QCD
while  the values of the couplings  $\alpha_F(r_1)$ and
$\alpha_F(r_0)$ in BPT are by $\sim 30\% (n_f =3)$ and $50\%
(n_f=0)$ larger  than  corresponding  lattice couplings. With the
use of the BPT potential good description of the bottomonium
spectrum is obtained.
\end{abstract}
\large

\section{Introduction}

The static $Q\bar Q$ interaction plays a special role in hadron physics and the
modern understanding of the  spin-independent part of the static potential
$V(r)$  comes from different approaches: QCD phenomenology \cite{1} - \cite{3},
perturbative  QCD \cite{4,5}, the analytical perturbation theory \cite{6}, the
background perturbation theory (BPT) \cite{7,8}, and lattice QCD
\cite{9}-\cite{11}. It  has been established that in the region $r\ga 0.2$ fm
this  potential can be taken as a sum of nonperturbative (NP) confining
potential plus a gluon-exchange term: \be V(r) =V_{NP} (r) + V_{GE}
(r).\label{1}\ee Such an additive form was   introduced by the Cornell group
already 30 years ago, just after discovery of charmonium \cite{1}, and later it
has been supported by numerous calculations in QCD phenomenology and also on
the lattice, where  for the lattice potential the parametrization like    the
Cornell potential  is used [9,10], \be
 V_{lat}(r) =\sigma r-\frac{e_{lat}}{r},~~ (r\ga 0.2 {~\rm fm})\label{2}\ee
 with $e_{lat}=const, ~~ \alpha_{lat}=\frac34 e_{lat}$.
(For the lattice static potential we  use  here the notation
 $\alpha_{lat} (e_{lat}$) to distinguish it from the
 phenomenological case).

 In BPT the additive form (\ref{1}) can  automatically be  obtained
 in the lowest approximation when both NP and  Coulomb terms
 satisfy the Coulomb scaling  law with an accuracy  a
few  percents, in good agreement with the lattice data
 \cite{12}.

 The explicit form of NP term is well  known now: $V_{NP} (r)
 =\sigma r$ at the $Q\bar Q$ separations $r> T_g$, where $T_g$ is
 the gluonic correlation length, $T_g\cong 0.2$ fm,  measured on the lattice \cite{13}. Such
linear behavior is valid up to the distances $r\sim 1.0$ fm, while
at larger $r$,due to the $q\bar q$--pair creation,  the
flattening of confining potential, is expected  to take place
[14]. Here in our paper we  restrict our consideration to the
region $r\la 1.0$ fm.

Numerous calculations of the meson spectra have shown that linear
potential defines gross features of the spectra, in particular,
the slope and  the intercept of  Regge -- trajectories for light
mesons [15]. At the same time the splittings between low-lying
levels, the fine-structure splittings, and the wave function at
the origin   in heavy  quarkonia  are shown  to be very sensitive
to the gluon-exchange part of the static potential (\ref{1})
\cite{2,16}: \be V(r) =\sigma r + V_{GE}(r),\label{3}\ee $$
V_{GE}(r) \equiv -\frac43 \frac{\alpha_{st}(r)}{r}.$$
Unfortunately, our knowledge of this term (or the vector coupling
$\alpha_{st}(r)$) in IR region remains  insufficient   and the
choice of $\alpha_{st}(r)$ essentially differs  in different
theoretical approaches. The  only common feature of
$\alpha_{st}(r)$, used in QCD phenomenology, BPT,  and  supported
by   the lattice calculations, is that this coupling freezes at
large distances, \be \alpha_{st} ({\rm~~ large~~}r) = const =
\alpha_{crit}.\label{4}\ee  but the true  value of $\alpha_{crit}$
is not fixed  up to now.
 For example, in
the Cornell potential, used in the phenomenology, the typical
values of the  Coulomb constant  in (\ref{2}) lie in the range
0.39$\div 0.45$ [1-3,16] while on the lattice in the same
parametrization (\ref{2}) $\alpha_{lat}$ has the value by 50$\div
30$\% smaller: in quenched approximation $(n_f=0)~~
\alpha_{lat}\cong 0.20 \div 0.23$  \cite{9}-\cite{11} and in
$(2+1)$ QCD  $ \alpha_{lat} \cong 0.30$  \cite{11}.

Even larger freezing value is obtained  if the asymptotic freedom
(AF)  behavior of $\alpha_{st}(r)$ (or $\alpha_{st}(q)$ in
momentum space) is taken into account. In BPT large
$\alpha_{crit}=0.58\pm 0.02$ follows from the analysis of the
splittings between low-lying levels in bottomonium [16,17] and
this number is in striking agreement with $\alpha_{crit}$,
introduced by Godfrey, Isgur  in [3] in  phenomenological way.
From the analysis of the hadronic decays of the $\tau$-lepton the
number $\alpha_s(1 GeV) \cong 0.9 \pm 0.1$ was  determined in Ref.
[18], while essentially smaller value, $\alpha_s (1 GeV)\cong
0.45$, was obtained  in  perturbative QCD  with higher order
corrections [19]. The Richardson potential as well as  "the
analytical perturbation theory" give even larger~~ $\alpha_{crit}
(n_f=3) =1.4$ \cite{6}. Thus  at present the  true freezing value
of the vector coupling, as well as $\alpha_s$(1 GeV), remains
unknown.

Besides  the vector coupling $\alpha_{st} (r) $ (in the static
potential)  the coupling $\alpha_F (r)$, associated with the
static force,    can be introduced,
\be
F(r) =V'(r) =\sigma + V'_{GE} (r) , \label{5}\ee
 \be V'_{GE} (r) \equiv \frac43 \frac{\alpha_F(r)}{r^2},
 \label{6}\ee  where by definition $\alpha_F(r)$ is
 expressed through
   $\alpha_{st} (r)$,  \be \alpha_F (r) =
  \alpha_{st} (r) - r \alpha'_{st} (r). \label{7}\ee
Then   an  important  information about   the derivative
$\alpha'_{st} (r)$ can be obtained from the study of the  force.
 Note that the matrix elements over the static force define the  squared wave function  at the origin for  the $nS$ states in heavy quarkonia.
 In the lattice analysis  the vector coupling $\alpha_{lat} $
  in $V_{lat}(r)$ (\ref{2}) is supposed to be independent of the
  $Q\bar Q$ separation $r$ over the whole region $0.2$ fm $\div 1.0$ fm
  [10], i.e.   $\alpha'_{lat} =0$,  and  therefore in this region $\alpha_{lat} \cong \alpha_F =const$.
  This statement   is in accord with   recent  calculations  of the lattice force in [11] where
  equal values  $\alpha_F^{lat} (r_1)=\alpha_F^{lat} (r_0)$  at
  the points $r_1\approx 0.34$ fm,  $r_0=0.5$ fm  have been  obtained.

   Different  picture  takes place  in BPT where the vector
   coupling $\alpha_B(r)$ at the points $r_1=0.34$ fm and
   $r_0=0.5$ fm changes by 20\% (see Section 4) being close to the
   freezing value only at large distances, $r\ga 0.6 $ fm.
 Also due to   the $r$-dependence   an essential difference between $\alpha_F(r)$ and
 $\alpha_{B}(r)$  occurs  just in the range $0.2 $fm $\leq r \la
 0.6$ fm.

 Therefore   several  features of the   vector coupling still need to be  clarified.

 First, why in lattice QCD and in QCD phenomenology the Coulomb
 constant, used in the  same  Cornell
 potential, differs by 30\% $(n_f=3)$?

 Second, whether   the $r$-dependence of the vector coupling
 $\alpha_{st} (r)$ (also of  $\alpha_F(r)$) really exists  and  why it is not
 observed on the lattice?

Third, what is the  true freezing value?

To answer some of  these questions we shall use  in our analysis
here BPT which gives a consistent analytical description of the
vector coupling both in momentum and in coordinate spaces. To test
the BPT conception about the vector-coupling behavior in the IR
region,
 recently the heavy-quarkonia spectra
  have been  successfully
described  in this approach with the use of only fundamental
quantities: the current (pole) quark mass,
$\Lambda_{\overline{MS}}(n_f)$, and the string tension [16],[17].
It is important  that in  BPT the vector coupling $\alpha_B(q)$
 has  correct perturbative limit at large $q^2$, and  therefore  it is
fully defined by the QCD constant $\Lambda_{\overline{MS}}$ and
also by so-called background  mass $M_B$ which is proportional $
\sqrt{\sigma}$.

Additional information about $Q\bar Q$ static interaction can
 be extracted   from the study of the static force $F_B(r)$ in
BPT, with further  comparison  to  recent lattice results  from  [11]. To this end it is convenient to calculate  the
dimentionless function $\Phi(r) = r^2 F(r)$  at  two
characteristic points -- $r_1\sqrt{\sigma},~~ r_0\sqrt{\sigma}$,
where  \be \Phi(r_1 \sqrt{\sigma}) =1.0 ~~ {\rm and ~~}
\Phi(r_0\sqrt{\sigma})=1.65.\label{8}\ee  On  the lattice the
values $\sqrt{\sigma} r_1^{(l)}, \sqrt{\sigma}r_0^{(l)}$ are shown
to be  different in quenched approximation and in  $(2+1)$
QCD. The same function $\Phi_B(r)$,  calculated here  in BPT,  aquires
the values  (8) at the  points  $\sqrt{\sigma} r_1$, $
\sqrt{\sigma}r_0$ which
 appear to be  only by  8\% smaller  than those on the lattice.
 However,  the freezing value $\alpha_{crit}$  in BPT is shown to be
 essentially larger than  $\alpha_{lat}$.
 To compare the BPT   and the lattice  potentials
  we calculate here the splittings between low-lying states in bottomonium and demonstrate that
   the lattice
 static potential cannot provide good agreement with experiment,
  while in BPT such an agreement takes place  for the conventional values of the pole mass of $b$ quark,
 the string tension, and $\Lambda_{\overline{MS}}(n_f=5)$.
We give also explanation why between the freezing values in BPT
and in the  Cornell potential, used in phenomenology, there is the
difference about 25\%.

  The paper is organized as follows. In Section 2 we shortly
  present the vector coupling in BPT and discuss the  correct choice
  of the QCD constant $\Lambda_V$ in the Vector-scheme. In Section
  3 the  characteristics of the  lattice force are presented
  while in Section 4 the values $r_1\sqrt{\sigma},r_0
  \sqrt{\sigma}$ are calculated in BPT. The difference between the
  vector couplings in both  approaches is also discussed. In
Section 5 the splittings between low-lying states in bottomonium
  are used as a test to compare the static potentials from the
  lattice  data, in phenomenology, and in BPT.      We show that
  the  lattice static potential cannot provide good agreement with
  experiment.  In Section 6 we
  present our Conclusion.

 \section{The strong coupling  in BPT}

 In BPT the gluon-exchange term
$ V_{GE}^B(r)$ defines the vector (background) coupling $\alpha_B
 (r)$ in the same way, as "the exact coupling"
$\alpha_{st} (r)$  is defined in Eq. (3):
\be
V_B(r)=\sigma r + V_{GE}^B(r);~~V_{GE}^B(r) =-\frac43
\frac{\alpha_B(r)}{r}\label{11}\ee With the use of the Fourier
transform of the  potential $V_{GE}^B(q)$ the background coupling
in coordinate space  can be expressed through the coupling
$\alpha_B(q)$ in momentum space [20]:
\be
\alpha_B(r) =\frac{2}{\pi} \int^\infty_0\frac{dq}{q}\sin (qr)
\alpha_B(q),\label{12}\ee where the vector coupling in momentum
space is defined at all momenta in Euclidean space and has no
singularity for $q^2>0$ [7]. In two-loop approximation  the
coupling  \be \alpha_B(q) =\frac{4\pi}{\beta_0 t_B}\left( 1-
\frac{\beta_1}{\beta^2_0} \frac{\ln t_B}{t_B}\right),\label{13}\ee
\be
t_B=\ln \frac{q^2+M^2_B}{\Lambda^2_V},\label{14}\ee contains the
background mass $M_B$  which enters under logarithm
 as a moderator of the IR behavior of the perturbative coupling. In
$\alpha_B(q)$ the Landau ghost pole disappears while the short
distance perturbative behavior, as well as the Casimir scaling
property of the static potential, stays intact [12].

The value of $M_B$ is determined by the lowest excitation  of a
hybrid [21], however this mass cannot be considered as an
additional (fitting) parameter in the theory, since in QCD string
theory it can be calculated on the same grounds  as mesons,  being
expressed through the only (besides $\Lambda_V)$ dimensional
parameter $\sqrt{\sigma}$: \be M_B=\xi \sqrt{\sigma}\label{15}\ee

We suppose here that in the static limit the coefficient $\xi$
does not depend (or weakly depends) on the number of flavors
$n_f$.  Direct calculation of $M_B$  has not yet   done and
therefore the number $\xi$ has been extracted from two fits: from
the comparison of lattice static potential to that in BPT [20] and
from the analysis of the spectra in charmonium and bottomonium
[16,17],  with the following result: \be M_B=2.236 (11)
\sqrt{\sigma}.\label{16}\ee
 In particular, the values $M_B =1.0$ GeV and 0.95 GeV  correspond to  $\sigma=0.20
 $ GeV$^2$ and 0.18 GeV$^2$.

 Note that the logarithm (\ref{14}) in $\alpha_B(q)$ formally
 coincides with that suggested in Refs. [22] many years ago in the
 picture where the gluon acquires an effective  mass $m_g$ and, as
 a result, in (\ref{13}) instead of $M^2_B$ the value $(2m_g)^2$
 enters. However, the physical gluon has no  mass and  in BPT the
 parameter $M_B$  has been interpreted in  correct  way as a hybrid
 excitation  of the $Q\bar Q$ string, which is proportional $\sqrt{\sigma}$ [21].

 The background coupling $\alpha_B(q)$ has correct PQCD limit at
 $q^2\gg M_B^2$ and therefore the constant $\Lambda_V(n_f)$ (in
 Vector-scheme)  under the logarithm (\ref{14}) can be expressed
 through the conventional QCD constant
 $\Lambda_{\overline{MS}}(n_f)$ as in PQCD [23]:
 \be \Lambda_V (n_f)=\Lambda_{\overline{MS}}(n_f) \exp
 \left(\frac{a_1}{2\beta_0}\right)\label{17}\ee
 with  $a_1=\frac{31}{3} -\frac{10}{9} n_f,~~ \beta_0 =11-\frac23
 n_f$. At present  the values of $\Lambda_{\overline{MS}}^{(n_f)}$
 are well established in two cases -- from
 high-energy processes for $n_f=5$ [24] and in quenched approximation
 from lattice calculations [25]:
 \be
 \Lambda^{(5)}_{\overline{MS}} (2-loop ) = (216\pm 25) {\rm
 ~MeV~}, \label{18}\ee
 which corresponds to the "world average" $\alpha_s(M_Z)=0.117\pm
 0.002$, and the value  \be
 \Lambda_{\overline{MS}}^{(0)} (2- loop) =\frac{0.602 (48)}{r_0}
 =(237\pm 19) ~{\rm MeV}\label{19}\ee
(with $r_0 =0.5$ fm $=2.538$ GeV$^{-1}$)  was calculated on the
lattice  in [25].

  Then from (\ref{17}) the corresponding
  values of $\Lambda_V{(n_f)}$  in the Vector-scheme are
  following,
  \be
  \Lambda_V^{(5)} = (295\pm 35)~{\rm MeV},\label{20}\ee
 \be
  \Lambda_V^{(0)} = (379\pm 30)~{\rm MeV}.\label{21}\ee
It is worthwhile to notice that the background coupling
$\alpha_B(r)$ (12) is  a universal function of the ratio \be
\lambda(n_f) =\frac{\Lambda_V(n_f)}{M_B}, \label{22}\ee  and
actually depends on  the dimensionless variable  $x=rM_B$ and
$\lambda$: \be \alpha_B(r, \Lambda_V, M_B)\equiv \alpha_B (rM_B;
\lambda)= \frac{2}{\pi} \int \frac{dx}{x} \sin (x, rM_B)
\alpha_B(x, \lambda). \label{23}\ee The  same ratio $\lambda$ also
defines the freezing (critical) value of the coupling (which
coincides
  in momentum and coordinate spaces), and  it is  given by the
analytical expression,
\be
\alpha_{crit} (q\to 0) =\alpha_{crit} (r\to\infty)
=\frac{4\pi}{\beta_0 t_{crit}} \left\{ 1-
\frac{\beta_1}{\beta_0^2} \frac{\ln
t_{crit}}{t_{crit}}\right\},\label{24}\ee
\be
t_{crit} =\ln \frac{M_B^2}{\Lambda^2_V} =\ln \lambda^2
(n_f).\label{25}\ee Taking  $\Lambda_V$ from (\ref{20}),
(\ref{21}) and $M_B$ (14) one obtains the following numbers for
$\alpha_{crit}$:

in
quenched approximation
\be
 \alpha_{crit} (n_f=0) = 0.419\begin{array}{l}+0.045\\-0.038,
 \end{array}\label{26}\ee
while  for $n_f=5$ the   freezing value is by $ \sim 30\%$ larger,
\be
 \alpha_{crit} (n_f=5) = 0.510\begin{array}{l}+0.055\\-0.049, ~~
\end{array}(M_B=1.0{~\rm GeV~})\label{27}\ee
and
\be
 \alpha_{crit} (n_f=5) = 0.533\begin{array}{l}+0.062\\-0.053, ~~
\end{array}(M_B=0.95{~\rm GeV~}).\label{28}\ee

 Note that   the freezing value (\ref{28})  turns out to be
very close to that, phenomenologically introduced  by Godfrey ,
Isgur  in [3] to describe a lot of experimental data in meson
sector. As shown in [17], the choice with $M_B=0.95$ GeV
$\Lambda_V^{(5)}\cong 320$ MeV, and $\alpha_{crit}=0.58$, provides
good agreement between experiment and BPT calculations of the
splittings in bottomonium. Thus, in BPT the   large freezing value
$\alpha_{crit} (n_f =5) \approx 0.58$ appears to be consistent
with the conventional value of
$\Lambda_{\overline{MS}}^{(5)}\approx 230 $ MeV, $\alpha_s (M_Z,
2-loop)=0.119 \pm 0.001$.

However,  the  large  freezing value in BPT  (and in QCD
phenomenology)  does not agree with the  value  used in the
lattice parametrization (2) of the static potential. For example,
 in quenched approximation $\alpha_{lat} \cong 0.23$ was obtained
 in  [9,10], while in BPT the \underline{minimal} value
 $\alpha_{crit} $ (which corresponds to the minimal value in
 (\ref{21}), $\Lambda_{\overline{MS}}^{(0)} (\min) =218$ MeV) is
 equal
 $\alpha_{crit}^{\min} (n_f=0) =0.38 $, i.e.,
 by 40\% larger. Such a difference between two  numbers   partly occurs due to the fact
 that on the  lattice the $r$-dependence  of the vector coupling is not seen (or
 neglected)   at $r>0.2$ fm .

 In Table 1  the background coupling $\alpha_B(r)$  is compared to
  $\alpha_{st}^P(r)$, calculated in PQCD, where according
 to the  perturbative prescription the QCD constant $\Lambda_R^{(n_f)} $
 in coordinate space is defined as $\Lambda_R^{(n_f)} =e^\gamma
 \Lambda_V^{(n_f)}$ ($\gamma$ is the Euler constant). These couplings (both in two-loop approximation) appear
 to be close to each other only at very  small distances, $r<0.1$ fm,
 and have the  characteristic  feature that the  difference $\Delta
 \alpha = \alpha_B(r) -\alpha_{st}^P(r)$ is positive at $r\leq
 0. 04$ fm and becoming negative at $r>0.04$ fm; just such a change
 of the sign of $\Delta \alpha$  has  been  observed  in lattice static
 potential [10].

\vspace{0.5cm}

\begin{tabular}{lp{11cm}}
{\bf Table 1}:& The strong coupling $\alpha^P_{st} (2-loop, r)$ in
PQCD and the background coupling $\alpha_B(2-loop, r)$ in BPT in
quenched approximation with $\Lambda^{(0)}_{\overline{MS}}=237$
MeV ($\Lambda_V^{(0)} =379$ MeV, $\Lambda_R^{(0)}=675$ MeV,
$M_B=1.0$ GeV)

\end{tabular}

\begin{center}
\begin{tabular}{|c|c|c|}\hline
&&\\
$r$(fm) & $\alpha_{st}^P(r)$ & $\alpha_B(r)$ \\ &&\\\hline
  0.01 & 0.128 & 0.138 \\
  0.02 & 0.156 & 0.166 \\
  0.04 & 0.202 & 0.204 \\
  0.06 & 0.248 & 0.232 \\
  0.08 & 0.301 & 0.254 \\
  0.10 & 0.368 & 0.272 \\
  0.12 & 0.457 & 0.288 \\
  0.14 & 0.588 & 0.301 \\ \hline
\end{tabular}
\end{center}

\section{The static force on  the lattice}

The study of the static force  is especially convenient through
the dimensionless function $\Phi_{lat} (r)$ since this function
depends on the dimensionless variable like $\frac{r}{a}$, where
$a$ is a lattice spacing , or on $r\sqrt{\sigma}$ (if the string
tension is taken as the only mass scale). Recently $\Phi_{lat}
(r)$ was measured by the MILC group both in quenched case and in
$(2+1)$ lattice QCD [11]
 with  the following results:

First, the function $\Phi_{lat}(r)= r^2 F_{lat}(r)
=(r\sqrt{\sigma})^2 F(r\sqrt{\sigma})$ acquires the value
$\Phi_{lat} (r_1^l)=1.0$ at the $Q\bar Q$ separation $r_1^l$: \be
r^l_1 \sqrt{\sigma} =0.769\pm 0.002~~ (n_f=3)\label{29}\ee
\be
r^l_1 \sqrt{\sigma} =0.833\pm 0.002 ~~(n_f=0).\label{30}\ee

Second, the function $\Phi_{lat}$ has the value $\Phi_{lat}
(r_0\sqrt{\sigma})=1.65$ at the following separation $r_0^l$ (the
Sommer scale):
\be
r_0 \sqrt{\sigma} =1.114\pm 0.002 ~~(n_f=3)\label{31}\ee
\be
r_0 \sqrt{\sigma} =1.160\pm 0.002 ~~(n_f=0).\label{32}\ee

Also in [11]  the ratio \be \frac{r_0^l}{r_1^l} =1.449(5)~~
(n_f=3)\label{33}\ee is defined with  a precision accuracy. If in
(27)-(31) one takes  the string tension $\sigma=0.20$ GeV$^2$,
then the  characteristic points are  $r_1^l =0.34$ fm,  $
r_0^l=0.49$ fm for $n_f=3$ and slightly larger,  $r^l_1=0.37$ fm,
$r^l_0 =0.51$ fm in quenched case. The  numbers  obtained can be
used to extract the vector coupling $\alpha_F^{lat}(r)$,
associated with the force in lattice QCD:
\be
F_{lat}(r) =\sigma +\frac{4}{3r^2} \alpha^{lat}_F
(r),\label{34}\ee
\be
\Phi_{lat}(r) =\sigma r^2 +\frac43 \alpha^{lat}_F(r).
\label{35}\ee From  (27)-(31) it follows that  the values of
$\alpha_F^{lat} (r)$  are equal at the points $r^l_1$ and $r_0^l$
 with a good accuracy: \be\alpha_F^{lat}
(r_1^{(l)})=\alpha^{lat}_F(r_0^{(l)})=0.307 (4) ~~ ~~
(n_f=3),\label{36}\ee
 \be\alpha_F^{lat} (r_1^{(l)})=\alpha^{lat}_F(r_0^{(l)})=0.229
(3) ~~ ~~ (n_f=0).\label{37}\ee

Note that in quenched case the value (35) for $\alpha_F^{lat}(r)$
numerically coincides with the lattice coupling
$\alpha_{lat}=0.23$ in the \underline{static} potential (where
$\alpha_{lat}=const$  is assumed over the whole region $0.2$ fm
$\leq r \la 1.0$ fm).  Thus existing  lattice data are consistent
with the assumption that the derivative in this region is equal
zero, $\alpha'_{lat}(r)=0$.

  The lattice number (34) in full QCD appears to be     essentially
smaller than the coupling, used in BPT and also in QCD
phenomenology, in IR region. In order to make a conclusion, which
value
 provides better description of experimental data, in Section 5 as a test
 we shall
 calculate the bottomonium spectrum  with  the lattice  as well as
 with
 the BPT   static potentials.

 \section{ The static force in BPT}

In BPT  the static force  can be presented as in (5),
\be F_B(r)
=\sigma +\frac{4}{3r^2} \alpha_F(r),\label{38}\ee where the
 coupling $\alpha_F(r)$, associated with the force $F_B(r)$,  can
 be expressed
 through the known coupling $\alpha_B(r)$ (10) and its
derivative:\be \alpha_F(r) =\alpha_B(r) -r\alpha'_B
(r).\label{39}\ee Correspondingly, the dimensionless function
$\Phi_B(r)$ is  \be \Phi_B(r) =r^2 F_B(r)=(\sqrt{\sigma} r)^2
+\frac43 \alpha_F(r).\label{40}\ee

The coupling $\alpha_F(r)$ can be easily calculated from the
expression (\ref{12}) and due to negative contribution of the term
with the derivative  it  appears to be essentially smaller than
$\alpha_B(r)$ (in the region $r\la 0.6$ fm). The values of
$\alpha_F(r)$ for  $n_f=0$ and $n_f=3$ are given in Tables 2,3,
from which   the $r$-dependence of $\alpha_F(r)$ is explicitly
seen:

i) At the distances $r=0.2$ fm; 0.35 fm, and 0.50 fm the coupling
$\alpha_F(r) (n_f=3)$ is smaller  than $\alpha_B(r)$ (in the
static potential)  by 25\%, 18\%, and 12\%, respectively.

ii) The derivative $\alpha'_B(r)$ is larger for larger $n_f$ being
approximately proportional $\beta_0^{-1}$ ($\beta_0=11-\frac23
n_f$).

iii) At the distance $r\approx 0.2$ fm the coupling $\alpha_F(r,
n_f)$ in BPT coincides with  $\alpha_{lat}$ both for $n_f=3$ and
$n_f=0$  but at larger $r>0.2$ fm it manifests essential growth:
for $n_f=3$ $\alpha_F(r=0.335$ fm) =0.376, $\alpha_F(r_0=0.493$
fm) =0.430, being by $\sim20\%$ and $\sim 40\%$ larger than
$\alpha_{lat} =0.306$ from [11].

\vspace{0.5cm}

\begin{tabular}{lp{11cm}}
{\bf Table 2}:& The  background  couplings $\alpha_{F}(r) =
\alpha_B(r) -r\alpha'_B(r)$ and $\alpha_B (r)$  in quenched
approximation ($\Lambda^{(0)}_V=379$ MeV, $M_B=1.0$ GeV,
$\alpha_{crit} =0.419$)

\end{tabular}

\vspace{0.5cm}
\begin{center}

\begin{tabular}{|c|c|c|c|c|c|}\hline
  $r$(fm) & $\alpha_F(r)$ &$\alpha_B(r)$ & $r$(fm)  & $\alpha_F(r)$ & $\alpha_B(r)$
  \\\hline
    0.099 & 0.188 & 0.272 & 0.355 & 0.312 & 0.379 \\
  0.118 & 0.201 & 0.289 & 0.394 & 0.324 & 0.386 \\
  0.138 & 0.213 & 0.301 & 0.433 & 0.335 & 0.391 \\
  0.158 & 0.225 & 0.313 & 0.473 & 0.345 & 0.396 \\
  0.197 & 0.246 & 0.333 & 0.493 & 0.350 & 0.398 \\
  0.236 & 0.265 & 0.348 & 0.532 &0.358 & 0.401 \\
  0.296 & 0.290 & 0.365 & 0.591 & 0.369 & 0.406 \\
  0.335 & 0.305& 0.375 & $\alpha_{crtit}$ & 0.419 & 0.419 \\ \hline
\end{tabular}
\end{center}

\begin{tabular}{lp{11cm}}
{\bf Table 3}:& The  background  couplings
$\alpha_{F}(r)=\alpha_B(r)-r\alpha'_B(r)$ and $\alpha_B (r)$ for
$n_f=3$  ($\Lambda^{(3)}_V=370$ MeV, $M_B=1.0$ GeV, $\alpha_{crit}
=0.510$)

\end{tabular}

\vspace{0.5cm}

\begin{center}

\begin{tabular}{|c|c|c|c|c|c|}\hline
  $r$(fm) & $\alpha_F(r)$ &$\alpha_B(r)$ & $r$(fm)  & $\alpha_F(r)$ & $\alpha_B(r)$
  \\\hline
  0.099 & 0.233 & 0.336 & 0.355 & 0.385 & 0.464 \\
  0.118 & 0.250 & 0.355 & 0.394 & 0.399 & 0.472 \\
  0.138 & 0.265 & 0.371 & 0.433 & 0.413 & 0.478 \\
  0.158 & 0.279 & 0.385 & 0.473 & 0.425 & 0.484 \\
  0.197 & 0.305 & 0.409 & 0.493 & 0.430 & 0.486 \\
  0.236 & 0.328 & 0.427 & 0.512 &0.435 & 0.488 \\
  0.296 & 0.358 & 0.448 & 0.532 & 0.440 & 0.490 \\
  0.335 & 0.376& 0.459 & 0.591 & 0.453 & 0.495 \\ \hline
\end{tabular}
\end{center}

 Knowing $\alpha_F(r)$  the function $\Phi_B(r)$ (\ref{40}) can be easily
calculated. Note that the coupling $\alpha_F(r)$ (as well as
$\alpha_B(r)$) weakly depends on the string tension through the
background mass  $M_B=2.236 \sqrt{\sigma}$ (16). We have obtained
the following numbers for the separations $r_1$ and $r_0$, where
$\Phi_B(r_1)=1.0$ and $\Phi_B(r_0)=1.65$:
\begin{eqnarray} r_1\sqrt{\sigma} &=&0.769(5)~~~~~~~~~~
(n_f=0)~\label{41a}\\r_0\sqrt{\sigma}& =& 1.090 (5),\nonumber
\end{eqnarray}   and the ratio $r_0/r_1$ coincides with
$r_0^{(l)}/r_1^{(l)} (n_f=0)$  on the lattice with 2\% accuracy:
$$ \frac{r_0}{r_1} =1.417 (16) (n_f=0). $$ For $n_f=3$ in BPT
 the characteristic separations  are following,
  \begin{eqnarray} r_1\sqrt{\sigma} &=&0.716 (4)\label{42}\\r_0\sqrt{\sigma}
&=&1.044 ~(5)~~ ~~~~~~~~(n_f=3)\nonumber \\ \frac{r_1}{r_0}
&=&1.458 (15)\nonumber \end{eqnarray}  The comparison of obtained
numbers to those  in lattice QCD shows that $r_1\sqrt{\sigma}
(r_0\sqrt{\sigma})$ in BPT is only by 8\% (6\%) smaller than  the
lattice values (27)-(30) while the ratio $r_1/r_0$ in (40)
coincides   with (31) with 1\% accuracy.

To have a full picture -- how the separations $r_1\sqrt{\sigma},
r_0\sqrt{\sigma}$ are changing  with  a increase  of flavors,
below we give their  values   also for $n_f=5$ (see Table 4):
\begin{eqnarray}
 \sqrt{\sigma
}r_1&=&0.673 (4), ~~~~~~~~~~~~(n_f=5)\label{43a} \\ \sqrt{\sigma
}r_0&=&1.016 (5), \nonumber \end{eqnarray}
 with their ratio
 $$ \frac{r_0}{r_1} =1.510 (16) ~~ (n_f=5).$$

\begin{tabular}{lp{11cm}}
{\bf Table 4}:& The  background  couplings $\alpha_{B}(r)$ and  $
\alpha_F(r)$  for  $n_f=5$ ($\Lambda^{(5)}_V=320$ MeV, $M_B=1.0$
GeV, $\alpha_{crit} =0.548$)
         \end{tabular}\\

\begin{center}

\begin{tabular}{|c|c|c|c|c|c|}\hline
 $r$(fm) & $\alpha_F(r)$ &$\alpha_B(r)$ & $r$(fm)  & $\alpha_F(r)$ & $\alpha_B(r)$
  \\\hline
 0.099    & 0.274 & 0.381 &  0.355  &0.435 &0.508 \\
 0.118  & 0.292 & 0.401 &  0.394 & 0.450&0.515 \\
 0.138  & 0.309 & 0.417 &  0.433 & 0.463 & 0.521 \\
 0.158  & 0.324 & 0.432 &  0.473  & 0.474 &0.526 \\
 0.197  & 0.352 & 0.455 &  0.493  & 0.479 & 0.528 \\
 0.236   & 0.377 & 0.473 &  0.512  & 0.484 & 0.530 \\
 0.296   & 0.409 & 0.494 &  0.532  & 0.488 & 0.532 \\
 0.335   & 0.427 & 0.504 & 0.591 & 0.501& 0.536 \\ \hline
\end{tabular}\\
\end{center}

  Comparing  (39)-(41) one can see  the points $r_1(n_f), r_0(n_f)$ are  smaller   for larger $n_f$
 (for $n_f=5$  they are  by 12\% \underline{smaller} than in
 quenched case).  Also   for $n_f=5$ the coupling
 $\alpha_F(r)$ approaches a freezing value at  smaller  distances, e.g. the value $\alpha_F(r_0)/\alpha_{crit}$
 is equal 0.835 $(n_f=0)$, 0.843 $(n_f=3)$, 0.874 $(n_f=5)$. (See also Fig.1.)

\vspace{1 cm}
\begin{figure}[!htb]
\includegraphics[width=.7\textwidth,angle=0]{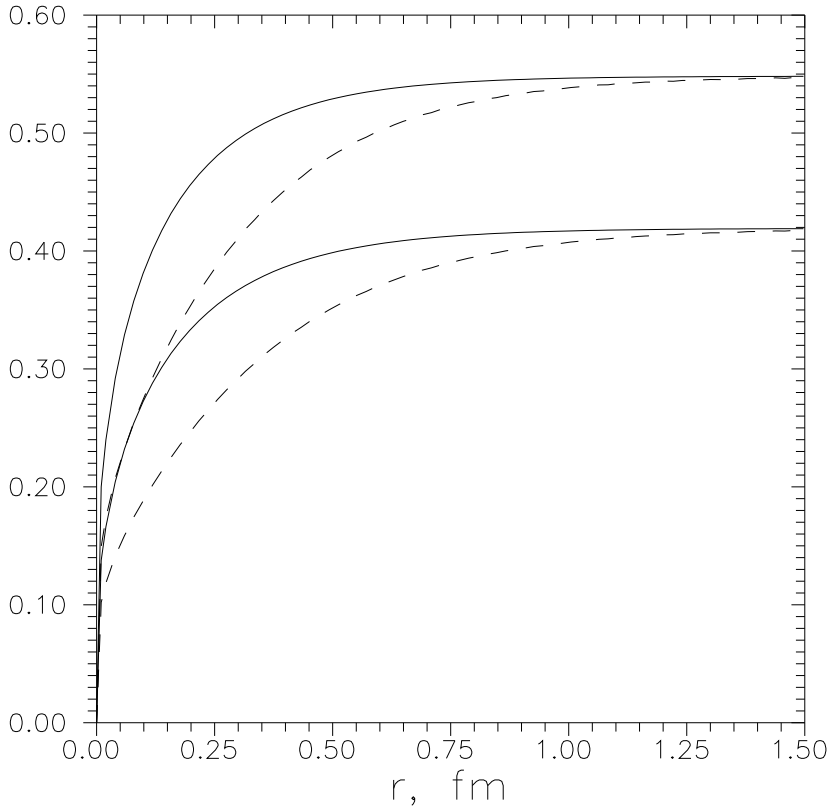}
 \caption{ The couplings $\alpha_B(r)$ solid lines)
 and  $\alpha_F(r)$ (dashed lines). The upper curves refer to case
 with $ n_f=5 $ and the lower curves refer to quenched case ($n_f=0$). }
\end{figure}

 In contrast to lattice coupling $\alpha^{lat}_F (r),$ which  is supposed to be
  $r$-- independent between $r_1$ and $r_0$, (34, 35),
  in BPT the coupling $\alpha_F(r)$ depends
 on $r$  and
  calculated  values of $\alpha_F(r_1),$  $\alpha_F(r_0)$ are given below:
 $$\alpha_F(r_1^{(0)}\sqrt{\sigma})=0.306~~ (n_f=0),$$
 \be \alpha_F(r_1^{(3)}\sqrt{\sigma})=0.366~~ (n_f=3),\label{43}\ee $$
\alpha_F(r_1^{(5)}\sqrt{\sigma})=0.412~~ (n_f=5),$$ and at the
Sommer scale $r_0\sqrt{\sigma}$ their values are by $\sim 12\%$
larger,  $$\alpha_F(r_0^{(0)}\sqrt{\sigma})=0.346~~ (n_f=0),$$
 $$\alpha_F(r_0^{(3)}\sqrt{\sigma})=0.42~~ (n_f=3),$$
\be\alpha_F(r_0^{(5)}\sqrt{\sigma})=0.46~~ (n_f=5).\label{44}\ee

From our analysis one can conclude that
\begin{description}
  \item[~~~i)]  The characteristic  quantities $r_1\sqrt{\sigma},
  r_0\sqrt{\sigma}$ in BPT are very close to the
  corresponding lattice values being only by 8\% smaller.
  \item[~~~ii)] At the same time the coupling
  $\alpha_F(r_0)$ in the "background force" $F_B(r)$  is shown  to be
by   30\%-40\% larger than $\alpha_F^{lat}$.
\end{description}

\section{The $b\bar b$ spectrum as a  test of  the vector
coupling in IR region}

Recently it was  demonstrated  that the bottomonium spectrum,
especially the splittings between low-lying levels, appear to be
very sensitive  both to  the  freezing  value  and   to the
$r$-dependence of the  vector  coupling  [16,17]. Therefore just
these splittings can be  used for  testing   of  $\alpha_{st}(r)$
in IR region. For illustration we
 consider here  three  typical  static  potentials.

First one imitates the  lattice  static potential  (2) but  has
the form  of   the Cornell potential  over the whole region $r\geq
0$. The coupling  $\alpha_{lat}(r)=const=0.306$  is taken from
lattice calculations with $n_f=3$ from [11] which is  the largest
Coulomb   constant obtained in lattice measurements up to now. In
this   potential  the AF behavior of $\alpha_{lat} (r)$ (which was
observed on the lattice at small $Q\bar Q$ separations, $r\la 0.2$
fm [9]) is neglected and  therefore   such the gluon-exchange
potential with $\alpha_{lat}=const$  over the whole region  is
stronger than the  lattice potential  and  gives  rise to  larger
values of the splittings between levels in bottomonium.

Second variant refers  to the phenomenological Cornell potential
with $\alpha_{st}(r)=const=0.42$ which is used in analysis of
charmonium [1,26].

Third potential is taken as in BPT for $n_f=5$, $\Lambda_V^{(5)} =
330$ MeV (or $\Lambda^{(5)}_{\overline{MS}}(2- {\rm loop}) = 241$
MeV), and $M_B =0.95$ GeV. The calculations with  first two
potentials are performed with nonrelativistic kinematics while for
third potential the solutions of  relativistic Salpeter equation
are used.

The splittings between the spin-averaged masses $\bar M(nL)$ in
bottomonium for these three potentials are given  in Table 5.\\

\begin{tabular}{lp{11cm}}
{\bf Table 5}:& The   splittings  (in  MeV) between  spin averaged
masses in  bottomonium  for the Cornell potential with   (A)
$\alpha_{lat} =0.306; (B) \alpha_{phen} =0.42$ and  for  the BPT
 potential  with $\Lambda_V^{(5)} =330$   MeV ($M_B=0.95$
GeV)

\end{tabular}

\vspace{0.5cm}

\begin{center}

\begin{tabular}{|c|c|c|c|c|}\hline
  Splitt.& A. $\alpha_{lat}=0.306$& B. $\alpha_{phen} =0.42$&
 C.  $\alpha_{crit}=0.565$&\\
& $ \sigma_A=0.20 $ GeV$^2$ &  $\sigma=0.183 $ GeV$^2$ &
$\sigma=0.178$ GeV$^2$& exper.\\ & $m_A=4.85$ GeV& $m_B=4.631$ GeV
& $m_b=4.828$ GeV&
  \\\hline
  2S-1S &  527 & 615 &557 & $563^{a)}$ \\
  2S-1P & 114 & 97 &123 & $ 123^{a)}$  \\
  1P-1S & 413 & 517 &434 &$440^{a)}$ \\
  1D-1P & 241 & 260 &260 &261$\pm2^{b)}$ \\
   2P-1P& 359& 368&370 & 360.1$\pm 1.2$\\\hline
\end{tabular}
\end{center}

$^{a)}$ Since the mass of $\eta_b(nS)$ is unknown and  in any case
$\bar M(1S)<M(\Upsilon (1S))$, we give here only the low limit of
the experimental  splitting.

$^{b)}$  The   experimental number for $M$(1D$_2) = 10161.2\pm
2.2$ is obtained in [27].

From our calculations presented in Table 5 one can make important
conclusions. First, the calculations with "the lattice" potential
$A$ with $\alpha_{lat}=0.306(n_f=3)$ define the upper bounds of
the splittings between different states in bottomonium (since the
AF behavior at $r\leq 0.2$ fm is neglected). Nevertheless even
upper bounds of the 2S-1S, 1P-1S splittings appear to be by
40$\div30$ MeV smaller than the experimental numbers (for which we
know the lower bounds since the $\eta_b(nS)$ mesons are still
unobserved,  $\bar M(nS) <M(\Upsilon (nS)))$.

It is of a special
 importance to compare theoretical and experimental number for the
 1D-1P splitting which is measured now with precision accuracy \cite{28}:
 \be
 \Delta (\exp) =\bar M ({\rm 1D}) -\bar M ({\rm 1P}) = 261.1 \pm 2.2 (\exp)
 \pm \begin{array}{l}1.0\\0.0 \end{array}(th) {\rm ~MeV},
 \label{44}\ee
where  $\bar M$(1P) =9900.1$\pm0.6$ MeV, $\bar
 M$(1D)$=M_{\exp}$(1D$_2)\pm \begin{array}{l}1.0\\0.0\end{array} (th)
 {\rm~MeV};$  $  M_{\exp} $(1D$_2)= 10161.2\pm 1.6 (\exp)$ MeV (see a discussion of  the 1D -1P splitting  in  \cite{17}).

 For "the lattice" potential A the upper limit  $\Delta(lat)$
 turns out to be by 20 MeV smaller than $\Delta(\exp)$ (the
 difference  is about ten standard deviations).

 On the contrary the calculations with the phenomenological
 potential $B$ (with $\alpha_{phen}=0.42$) and with the BPT
 potential $C$ give the  precision agreement with  $\Delta (\exp)$ .

In BPT the 2S-1S, 1P-1S splittings   are  also close to  the
 experimental numbers being  only by $\sim 10$ MeV
 smaller (the ground state mass  $\bar M$(1S)=9466
 MeV is   slightly larger than expected experimental number).
The same splittings for  the  Cornell potential
 ($\alpha_{phen}=const =0.42)$ turn to  be  too large
 (since  $\bar M$(1S)$\cong 9300$ MeV is small) because the
 Coulomb part of the static potential is overestimated if  the  AF
 behavior  of the vector coupling is neglected.  This AF
 effect  is small  for  the 1D-1P splitting for
 which agreement  with experiment takes place.

Note that in the BPT potential $C$ the freezing value
$\alpha_{crit}=0.565$ is essentially larger than
$\alpha_{phen}=0.42$ and to understand what kind of the
approximation  corresponds to $\alpha_{st} (r) =const$, let us
introduce an effective coupling in BPT according to the relation:
\be
\lan \frac{\alpha_B(r)}{r}\ran_{nL} = \alpha_{eff} (nL) \lan
\frac{1}{r}\ran_{nL}\label{45}\ee Our calculations of the matrix
elements demonstrate (see   Table 6) that

 (1) $ \alpha_{eff}$ depends on the quantum numbers $n,L$;

(2) the values of $\alpha_{eff}(nL)$ appear to be by 30$\div$15\%
smaller than the freezing value $\alpha_{crit} =0.565$ and those
values
 for the
1S, 2S states  are  close to $\alpha_{phen}$ used in
phenomenology.

\vspace{0.5cm}
\begin{tabular}{lp{11cm}}
{\bf Table 6}:& The    effective vector coupling  $\alpha_{eff}
(nL)$ for the BPT potential $C$ ~($\Lambda_V^{(5)} =330$ MeV,
$M_B=0.95$ GeV, $\sigma=0.178 $ GeV$^2, \alpha_{crit}=0.565$)\\

\end{tabular}

\begin{center}

\begin{tabular}{|c|c|c|c|c|c|c|c|}\hline
 state  & 1S & 2S & 3S & 1P & 1P & 1D & 2D \\\hline
$\alpha_{eff} (nL)$ & 0.405   & 0.439 & 0.448 & 0.495 & 0.501 &
0.528&0.528
\\ \hline
\end{tabular}
\end{center}
\vspace{0.5cm} (3)  For the orbital excitations the effective
coupling $\alpha_{eff} \cong 0.50$ is  by  $\sim 20\%$ larger than
for the 1S, 2S states and  just this fact results in increasing of
the splittings like 2S-1P, 1D-1P which is observed in bottomonium.

To make a decisive conclusion about  the behavior of
$\alpha_{st}(r)$ in  IR region it would be    also important  to
take into account a screening of the gluon-exchange potential at
large distances. At present there is no  a theory of this
phenomenon  on the fundamental level, although in  some cases the
Coulomb screening is introduced in a phenomenological way
\cite{29}.

\section{Conclusion}

Our study of the vector couplings $\alpha_B(r)$ (in the static
potential) and  $\alpha_F(r)$ in the static force $F(r)$ is
performed in the framework of BPT with the  following results.

Due to the derivative  term $\alpha'_{B}(r)$ an essential
difference between $\alpha_F(r)$ and $\alpha_B(r)$ is observed at
the $Q\bar Q$ separations $r\la 0.6$ fm with $\alpha_F(r)$  being
by 50\%, 30\%, 15\%  smaller    at  the  points $0.2$ fm, 0.3 fm,
and 0.5 fm, respectively.

At the same time the freezing values of both couplings coincide
 and  are   rather large: $\alpha_{crit} \cong 0.41(n_f=0)$; 0.51
$(n_f=3)$, and $\alpha_{crit}=0.58\pm 0.04$ for $n_f=5$.  The last
number turns out to be very close to that introduced by Godfrey,
Isgur in their phenomenological analysis.

 The dimensionless quantities  $r_1\sqrt{\sigma}$ and
 $r_0\sqrt{\sigma}$, where the function $\Phi(r)=r^2F(r)$ has the
 values 1.0 and 1.65, are calculated  in BPT  and their values
  are   by
 (6$\div$8)\% smaller that those  calculated on the   lattice
  in quenched case and in (2+1) QCD.

 In contrast to lattice observation  where
 $\alpha_F^{lat}(r_1)=\alpha_F^{lat}(r_0)=const$   and this  constant is
 small: $\alpha_{lat}=0.23(n_f=0)$ and $\alpha_{lat}
 =0.306(n_f=3)$, in BPT    $\alpha_F(r)$  at these points is found to
 be by 40\% larger for $n_f=0$ and by 30\%  larger for $n_f=3$.
 Because the Coulomb constant in the lattice static  potential is
 small, this potential gives essentially smaller 2S-1S, 1P-1S,
 1D-1P splittings in bottomonium.

The meaning of the Coulomb constant $\alpha_{phen}$, used in
phenomenology, as an effective coupling in BPT
 is suggested.  This interpretation   explains why  $\alpha_{phen}\cong
0.42$  is by 30\% smaller than the freezing value $\alpha_{crit}
\cong 0.56$  for  the potentials with the AF taken into
account.

The knowledge of the static force in BPT is important to perform
precision calculations of the wave functions at the origin,
hadronic decays, and fine structure splitting  in bottomonium.

\section{Acknowledgement}

For many years the authors  have had remarkable opportunity to
discuss a lot of problems in hadron physics with Prof.Simonov. His
scope and the attitude in physics had an enourmous influence on
us. We appreciate very much our collaboration with Yu.A.Simonov
and his permanent support in our research activity.

This work was partly supported by PRF grant for leading scientific schools No
1774.2003.2 .

 \end{document}